\documentstyle[preprint,pra,aps,eqsecnum]{revtex}

\begin{document}

\title{Linear stochastic wave-equations for continuously measured
 quantum systems}
\author{Peter Goetsch and Robert Graham}
\address{Universit\"at GH Essen, Fachbereich Physik, 45117 Essen,
 Germany}

\maketitle

\begin{abstract}
 While the linearity of the Schr\"odinger equation and the
 superposition principle are fundamental to quantum mechanics, so are
 the backaction of measurements and the resulting nonlinearity. It is
 remarkable, therefore, that the wave-equation of systems in
 continuous interaction with some reservoir, which may be a measuring
 device, can be cast into a linear form, even after the degrees of
 freedom of the reservoir have been eliminated.
 The superposition principle still holds for the stochastic
 wave-function of the observed system, and exact analytical solutions
 are possible in sufficiently simple cases. We discuss here the
 coupling to Markovian reservoirs appropriate for homodyne,
 heterodyne, and photon counting measurements. For these we present
 a derivation of the linear stochastic wave-equation from first
 principles and analyze its physical content.
\end{abstract}


\newpage
\section{Introduction}
Perhaps the most fundamental principle underlying quantum mechanics
is the superposition principle for quantum amplitudes which is
equivalent to the linearity of the Schr\"odinger equation. However,
as is well known, the time-evolution of the state vector $|\psi
\rangle$ of any system is governed by the linear Schr\"odinger
equation only as long as the system is decoupled from its
environment, and, in particular, from any extraneous measuring
apparatus. E.~g.~if during a short time interval such an extraneous
coupling is introduced, which performs a measurement of an observable
$X$ without destroying the system, the linear time-evolution is
interrupted; the wave-function is replaced by a new wave-function
$|\psi'\rangle$ chosen at random among the eigenfunctions
$|\psi_n\rangle$ satisfying $X|\psi_n\rangle=X_n|\psi_n\rangle$ with
the eigenvalues $X_n$; the probability for $|\psi_n\rangle$ is given
by
\begin{equation}
\label{1-1}
 p_n=|\langle\psi_n|\psi\rangle|^2\, ,\quad n\in {\rm I}
\end{equation}
where I is the set of all possible quantum numbers, here taken as
discrete, for simplicity. This step in the time-evolution is
therefore random and nonlinear. In order to formulate it explicitely
let us introduce a classical random variable $N$, which only takes
the values $N=n$, $n\in {\rm I}$ with probabilities
\begin{equation}
\label{1-2}
 P(N=n)\,=\,p_n\, ,\quad \sum_{n\in I}p_n=1.
\end{equation}
Then we may write for a measurement
\begin{equation}
\label{1-3}
 |\psi\rangle\rightarrow|\psi'\rangle=|\psi_N\rangle.
\end{equation}
If similar measurements are performed repeatedly in time, say at
the discrete time $t_i$, then the linear deterministic
time-evolution between successive measurements
is repeatedly interrupted by steps like (\ref{1-3}) with more random
variables $N_i$ whose probabilities follow again from eq.~(\ref{1-1})
and are therefore correlated, in general.

The possibility to describe measured systems, or more generally
systems coupled to their environment, by stochastic wave-functions
similar to (\ref{1-3}) is of course well known. By the work of many
authors over the years this possibility has been turned into an
efficient tool, useful in particular for numerical simulations, by
deriving equations of motion for the wave-functions of systems
continuously coupled to their environment. All of the early and
some of the more recent work in this area is purely phenomenological
(see e.g. \cite{1,2,3,4,5}) and often motivated by the desire to find
generalizations of the Schr\"odinger equation on a fundamental level
including wave-function collapse. Some of the more recent
work, however, stresses microscopic derivations, in certain weak
coupling and Markovian limits, from the Schr\"odinger equation for
the coupled system and reservoir (see e.g. \cite{8,9,7,11}). As one
may expect from eq.~(\ref{1-3}), the resulting wave-equations
generalize the Schr\"odinger equation for isolated systems in two
ways: they contain in an explicit way classical stochastic variables
analogous to $N$ in eq.~(\ref{1-3}), and they are nonlinear,
analogous to the nonlinear dependence of $N$ on $|\psi\rangle$ in
eq.~(\ref{1-3}) via its distribution (\ref{1-2}), (\ref{1-1}).

The appearance of stochastic elements in the generalized
Schr\"odinger equation may look unusual at first sight but should
not cause surprise, as it corresponds directly to the observed
stochastic behaviour e.~g.~of measured quantum systems. However,
the appearance of a nonlinearity in quantum measurements, even
though well known to all physicists, may still be considered as
alarming, because it seems to change the foundations of quantum
mechanics: In the real world, perhaps with the exception of the
universe as a whole, there aren't any systems which are completely
decoupled from their environment. This seems to lead to the alarming
conclusion that linearity and the superposition principle are only
approximate concepts for any system in the real world. Indeed, any
finite real-world system in an excited state will sooner or later
decay via one or several random nonlinear steps similar to
eq.~(\ref{1-3}). But if the nonlinearity of the environment-coupled
reduced time-evolution must be accepted, the role of the
superposition principle becomes obscure. E.~g.~it may no longer be
possible to analyze the fate of an initially given linear
superposition of states by studying the fate of each of its
constituents separately.

In view of this fundamental problem raised by the nonlinear
stochastic evolution (\ref{1-3}) it is very remarkable that the
stochastic nonlinear wave-equations, describing a broad class of
systems in continuous interaction with their environment, can
actually be cast into a linear form, containing the same information.
These linear equations still contain classical stochastic variables,
however now distributed independently of the wave-function. The noise
they describe is therefore independent of the system under study and
may be considered a property of the environment alone.

The existence of these linear versions of the stochastic
wave-equations has repeatedly been noted in the literature
\cite{7,11,6,10,14}, however their fundamental relevance in
connection with the superposition principle has, to our knowledge,
not been stressed. In the present paper we wish to derive and study
in a unified manner the linear stochastic wave-equations
corresponding to homodyne, heterodyne and photon-count measurements
in quantum optics.

These measurements have been considered before by Wiseman and Milburn
\cite{8,9} using nonlinear stochastic wave-equations. A microscopic
derivation of a linear stochastic wave-equation corresponding to
homodyne measurements was presented by Belavkin \cite{7,11}.
Our derivation starts from the same assumptions, but appears to be
simpler, at least to us. Photon-counting has been discussed in the
framework of phenomenological nonlinear and a phenomenological linear
stochastic wave-equations in papers by Barchielli \cite{14,13} and by
Belavkin \cite{6}. The linear stochastic wave-equation for
photon-counting, which we shall derive in the present paper is
different from the phenomenological version used in their work.
Stochastic Schr\"odinger equations with quantized noise terms have
been discussed in detail in the work of Gardiner et al \cite{Z}.
While these authors consider equations in the large Hilbert space of
system and environment, our work here is concerned with
wave-equations in the Hilbert space of the system without
environment. An extensive review of the use of nonlinear stochastic
wave-equations in quantum optics and a large list of references is
given in \cite{15}.

The derivation and use of stochastic wave-equations, both nonlinear
and linear, is so far restricted to Markovian reservoirs and we shall
also adopt this framework here. It remains an interesting open
problem for future work to examine also non-Markovian extensions.

\section{Linear stochastic description of a general quantum
measurement}
Here we shall present the general idea to be applied in the
subsequent sections for the case of a general quantum measurement.
It is our aim to show how the stochastic description to be applied
in subsequent sections is firmly rooted in the basic formalism
established by von Neumann (1931) \cite{16}, Wigner (1963) \cite{17}
and others.

The nonlinear stochastic change of a wave-function due to the
idealized instantaneous measurement of a quantum observable was
recalled in eq.~(\ref{1-3}). The nonlinearity resides in the
dependence of the probability $p(N=n)=|\langle\psi|\psi_n\rangle|^2$
on $|\psi\rangle$. We now turn to a stochastically equivalent linear
description, i.e. one which gives the same results for physical
probabilities.

We may begin by noting that the transformation (\ref{1-3}) would
trivially be linear if the distribution of the random variable $N$
would be independent of the state $|\psi\rangle$. Let us therefore
introduce a new classical random variable $\tilde{N}$ also
distributed over all quantum numbers $n\in {\rm I}$ with some fixed
but arbitrary distribution $\tilde{p}_n$ satisfying
\begin{equation}
\label{2-1}
  p(\tilde{N}=n)=\tilde{p}_n>0,\quad \tilde{p}_n\neq 0\,
  \forall\, n,\quad
   \sum_{n\in{\rm I}}\tilde{p}_n=1.
\end{equation}
Let us then consider the replacement of the nonlinear transformation
(\ref{1-3}) by the linear one
\begin{equation}
\label{2-2}
|\psi\rangle\rightarrow|\tilde{\psi}_{\tilde{N}}'\rangle=
 \frac{1}{\sqrt{\tilde{p}_{\tilde{N}}}}|\psi_{\tilde{N}}\rangle
\langle\psi_{\tilde{N}}|\psi\rangle
\end{equation}
where $\tilde{N}$ is distributed according to (\ref{2-1}). We note
that $|\tilde{\psi}'_{\tilde{N}}\rangle$ is not normalized,
\begin{equation}
\label{2-3}
\langle\tilde{\psi}'_{\tilde{N}}|\tilde{\psi}'_{\tilde{N}}\rangle=
\frac{1}{\tilde{p}_{\tilde{N}}}
|\langle\tilde{\psi}_{\tilde{N}}|\psi\rangle|^2\neq 1
\end{equation}
i.e. the transformation (\ref{2-2}) is not unitary. However, we have
instead
\begin{equation}
\label{2-4}
\sum_{n\in{\rm I}}\tilde{p}_n
 \langle\tilde{\psi}'_n|\tilde{\psi}'_n\rangle=1.
\end{equation}
The probability $P_X(X_n)$ to observe the eigenvalue $X_n$ of the
measured observable $X$ is fully encapsulated in the linear
stochastic transform $|\tilde{\psi}'_{\tilde{N}}\rangle$ of $|\psi
\rangle$ via the formula
\begin{equation}
\label{2-5}
P_X(X_n)=\tilde{p}_n
\langle\tilde{\psi}'_n|\tilde{\psi}'_n\rangle.
\end{equation}
Using eq.~(\ref{2-2}) in eq.~(\ref{2-5}) it can be seen that the
usual formula $P_X(X_n)=|\langle\psi_n|\psi\rangle|^2=p_n$ results
and, furthermore, that $\tilde{p}_n$ drops out completely and may
remain arbitrary, as long as the conditions (\ref{2-1}) are
satisfied. E.g. one may replace $\tilde{p}_n$ by a constant and
interpret the corresponding equidistribution of $\tilde{N}$ over the
set I as reflecting our complete ignorance of the outcome of the
measurement prior to the specification of $|\psi\rangle$ and prior to
the measurement. Correspondingly, eq.~(\ref{2-5}) represents the
transformation from the 'input' distribution $\tilde{p}_n$,
reflecting our ignorance, to the 'output'-distribution $p_n$ produced
by the measurement. This particular way of thinking about
eq.~(\ref{2-5}) will be useful lateron, where descriptions analogous
to eq.~(\ref{2-2}) are applied to various concrete measurements which
are performed continuously in time. In fact, concrete expressions for
the probability $\tilde{p}_{\tilde{N}}$ are obtained in these
concrete models.

{}From eq.~(\ref{2-5}) all moments of the measured observable may be
calculated. Furthermore, it is straightforward to check that for
measurements performed successively in time multi-time joint
probabilities of the observables and the correlation functions
associated to them may also be calculated by the repeated use of
eq.~(\ref{2-2}) for each measurement. Hence, eq.~(\ref{2-2}) provides
a description of the measurement of $X$ which is just as complete as
the nonlinear description (\ref{1-3}), but linear. E.g. if the
transformation (\ref{2-2}) is known for two wave-functions
$|\psi_A\rangle$ and $|\psi_B\rangle$ then we have for their linear
superposition
\begin{eqnarray}
\label{2-6}
C_A|\psi_A\rangle+C_B|\psi_B\rangle\rightarrow &&
         C_A|\psi'_A\rangle+C_B|\psi'_B\rangle\nonumber\\
    &&= \frac{1}{\sqrt{\tilde{p}_{\tilde{N}}}}|\psi_{\tilde{N}}
       \rangle\left(C_A\langle\psi_{\tilde{N}}|\psi_A\rangle+C_B
             \langle\psi_{\tilde{N}}|\psi_B\rangle\right).
\end{eqnarray}
We note that this expression still describes the stochastic
'collapse' of the superposition to a single state $|\psi_{\tilde{N}}
\rangle$. Yet the use of eq.~(\ref{2-6}) in the formula (\ref{2-5})
correctly predicts the interferences of the superposition observed
in the measurement of $X$ based on the behavior of the individual
components of the mixture.

We may ask now whether the wave-function $|\tilde{\psi}'_{\tilde{N}}
\rangle$ is physical, i.e. whether it corresponds in any sense to the
actual state of the reduced system. This question is difficult or
perhaps impossible to answer, even for the wave-function of closed
systems, which may well be considered as being devoid of physical
reality and no more than a mere computational tool to obtain physical
probabilities and their moments via some definite algorithm. The same
may be said for the wave-function $|\tilde{\psi}'_{\tilde{N}}\rangle$,
the only difference being that the usual algorithm for obtaining
probabilities is slightly changed according to eq.~(\ref{2-5}).

In order to derive eq.~(\ref{2-2}) from the Schr\"odinger equation
we may use the method of von Neumann (1931) \cite{16} and
Wigner (1963) \cite{17}. They show that an uncorrelated initial state
$|\psi\rangle$ of the measured system and $|\phi\rangle$ of the
measuring apparatus evolves according to
\begin{equation}
\label{2-7}
|\psi\rangle\otimes|\phi\rangle\rightarrow\sum_n\langle\psi_n
|\psi\rangle|\psi_n\rangle\otimes|\phi_n\rangle
\end{equation}
into an entangled state. The entanglement in eq.~(\ref{2-7}) is real
and could in principle be observed if further joint measurements on
system and apparatus are made. However, as long as further
measurements are made {\it only} on the system or {\it only} on the
apparatus the relative phases between the components of
eq.~(\ref{2-7}) for different $n$ are unobservable,
because both the $|\psi_n\rangle$ and the $|\phi_n\rangle$ are
orthogonal for different $n$, the $|\psi_n\rangle$ because they
correspond to different eigenvalues $X_n$, the $|\phi_n\rangle$
because they are macroscopically distinguishable states. Hence, as
far and only as far as such restricted further measurements are
concerned, a super-selection rule holds for $n$, (i.e. states of
different $n$ are not coherently superposed) and the right hand side
of eq.~(\ref{2-7}) is equivalent to a mixture
\begin{equation}
\label{2-8}
\sum_n|\langle\psi_n|\psi\rangle|^2|\psi_n\rangle
 \langle\psi_n|\otimes
  |\phi_n\rangle\langle\phi_n|.
\end{equation}
This restriction of the type of further measurements allows to
circumvent von Neumann's and Wigner's argument that a pure state
cannot evolve into a mixture, an old stumbling block in quantum
measurement theory. Tracing out the apparatus we end up with a mixed
state, which is equivalent to the description (\ref{2-2}) and
(\ref{2-1}) for arbitrary $\tilde{p}_n$.

\section{Homodyne measurements}
Let us briefly recall the principle of homodyne measurements. A
quantum oscillator or mode, say the electromagnetic mode of a cavity
at a given frequency but with (in comparison slowly varying)
time-dependent amplitude, radiates into the vacuum, where the
classical field of a local oscillator at the same frequency with
stable phase and time-independent amplitude is superimposed and the
beat intensity is detected. In balanced homodyning a difference
scheme is employed to get rid of the dominating field intensity of
the local oscillator in the output.

The input-output formalism \cite{18} is perfectly adapted to describe
this situation: The interaction of the system, i.e. the quantum
oscillator, with an input field generates an output field depending
linearly on the amplitude of the oscillator via a coupling constant.
In the case at hand the input field are just vacuum fluctuations
entering the quantum oscillator by the time-reverse of the channel
through which it radiates into the vacuum. For a differential
time-step we may write
\begin{equation}
\label{3-1}
dB_{\text{out}}(t)=\sqrt{\gamma}a(t)+dB_{\text{in}}
(t).
\end{equation}
Here $a(t)$, $a^\dagger(t)$ are the mode operators of the oscillator
in the interaction picture (i.e. with the oscillator frequency
$\omega_0$ splitt off); $\sqrt{\gamma}$ is a coupling constant which
we take positive by appropriately adjusting the origin of the phase
of $a$; $dB_{\text{in}}(t)$ represents the future directed
increments of the quantized input-field which is taken as quantized
Gaussian white noise \cite{19}. Therefore
\begin{eqnarray}
\label{3-2}
[dB_{\text{in}}(t),dB_{\text{in}}(t')] &=& 0 =
   [dB^\dagger_{\text{in}}(t),dB^\dagger_{\text{in}}
(t')]\nonumber\\~
[dB_{\text{in}}(t),dB^\dagger_{\text{in}}(t')] &=& 0 \text{ for } t
\neq t'\nonumber\\~
[dB_{\text{in}}(t),dB^\dagger_{\text{in}}(t)] &=& dt
\nonumber\\~
[dB_{\text{in}}(t),a(t')] &=&[dB^\dagger_{\text{in}}
(t),a(t')]=0 \text{ for } t\ge t'.
\end{eqnarray}
Acting on the vacuum we have $dB^\dagger_{\text{in}}(t)
dB_{\text{in}}(t)=0$. As the increments $dB_{\text{in}}
(t)$, $dB^\dagger_{\text{in}}(t)$ are causally disconnected
with the quantum oscillator at time $t$ all their commutors with the
oscillator variables at the same time vanish. The operator
$dB_{\text{out}}(t)$ in eq.~(\ref{3-1}) is the output field
generated in $dt$ by the interaction of the system oscillator with
the vacuum. The Hamiltonian for this interaction is
\begin{equation}
\label{3-3}
H_{\text{int}}(t)\,dt=H_0(a,a^\dagger,t)dt+i\sqrt{\gamma}
(adB_{\text{in}}^\dagger(t)-dB_{\text{in}}(t)
a^\dagger).
\end{equation}
Here $H_0(a,a^\dagger,t)$ is the part of the Hamiltonian responsible
for the generation of the quantum oscillation, in the interaction
picture. With the unitary operator
\begin{equation}
\label{3-4}
U_{\text{int}}(t)=\left[\exp\left(-i\int_0^t ds\,
H_{\text{int}}(s)\right)\right]_+
\end{equation}
where $[\dots]_+$ denotes time-ordering, we may then write
\begin{equation}
\label{3-5}
dB_{\text{out}}(t)=U_{\text{int}}^\dagger(t)dB_{\text{in}}(t)
U_{\text{int}}(t)
\end{equation}
which gives back eq.~(\ref{3-1}). The interaction Hamiltonian and
eq.~(\ref{3-1}) are therefore indeed consistent. As $dB_{\text{in}}
(t)$ and $dB_{\text{out}}(t)$ are connected by a
unitary transformation, the commutation relations of $dB_{\text{out}}
(t)$, $dB_{\text{out}}^\dagger(t)$ at equal time
coincide with those of $dB_{\text{in}}(t)$, $dB^\dagger_{\text{in}}
(t)$. The signal measured in the homodyne
experiment may now be expressed in terms of the output
field and its adjoint as
\begin{eqnarray}
\label{3-6}
|\beta|\Theta(t) &=& |\beta|\int_0^t d\Theta(s)\nonumber\\
|\beta|d\Theta(t) &\equiv&|\beta|\left(dB_{\text{out}}(t)
e^{-i\varphi}+dB^\dagger_{\text{out}}(t)e^{i\varphi}\right).
\end{eqnarray}
Here $\beta e^{i\varphi}$ is the complex amplitude of the classical
local oscillator. If this is expressed in terms of the input field
we find
\begin{equation}
\label{3-7}
d\Theta(t)=2\sqrt{\gamma}X(t)dt+d\Xi(t)
\end{equation}
with the input field
\begin{equation}
\label{3-8}
d\Xi(t)=(dB_{\text{in}}(t)e^{-i\varphi}+dB_{\text{in}}^\dagger(t)
e^{i\varphi})
\end{equation}
and the system observable
\begin{equation}
\label{3-9}
X(t)=\frac{1}{2}(ae^{-i\varphi}+a^\dagger e^{i\varphi})
\end{equation}
measured in this experiment.

The commutation relations (\ref{3-2}) imply that
\begin{equation}
\label{3-10}
\left[d\Xi(t),d\Xi(t')\right]=0
\end{equation}
and it follows from eq.~(\ref{3-7}) and eq.~(\ref{3-2}) that
\begin{equation}
\label{3-11}
\left[d\Theta(t),d\Xi(t)\right]=0=\left[d\Theta(t),X(t)\right].
\end{equation}
Therefore, there is a representation in which all operators in
eq.~(\ref{3-7}) are simultaneously diagonal and may be replaced by
their simultaneous eigenvalues, which we shall denote by
corresponding lower case letters. Thus
\begin{equation}
\label{3-12}
d\theta(t)=2 \sqrt{\gamma}x(t)dt+d\xi(t).
\end{equation}
The measured output field is therefore generated by the incoming
vacuum fluctuations and the instantaneous eigenvalue of the
$X$-quadrature component of the quantum oscillator. The relation
(\ref{3-12}) implies that the output process $d\theta(t)$, like the
input process obeys the rule of Ito calculus $(d\theta(t))^2=dt$.

Eq.~(\ref{3-12}) defines the output-process $d\theta(t)$ conditioned
on the input-process $d\xi(t)$ and the value of $x$ at time
$t$. In particular, if the distributions of $d\xi(t)$,
$p_\xi(d\xi(t))$, and of $x$ at $t$ are independent, as will
turn out to be the case lateron, the distribution of $d\theta(t)$
conditioned on $x(t)$ is given by
\begin{equation}
\label{Y}
p_\theta\left(d\theta(t))=p_\xi(d\theta(t)-2\sqrt{\gamma}x(t)dt
\right)
\end{equation}

It remains to derive the stochastic Schr\"odinger equation. The
generator of the total time-evolution follows from
\begin{equation}
\label{3-13}
d|\psi(t)\rangle=dU_{\text{int}}(t)|\psi(0)\rangle=
dU_{\text{int}}(t)U^\dagger_{\text{int}}(t)|\psi(t)
\rangle
\end{equation}
as
\begin{equation}
\label{3-14}
L(t)dt=dU_{\text{int}}(t) U^\dagger_{\text{int}}(t)
\end{equation}
where $dU_{\text{int}}(t)$ is calculated from $U_{\text{int}}(t)$
by using the Ito calculus for the stochastic
differential $dB_{\text{in}}(t)$ and its adjoint. The result
is
\begin{eqnarray}
\label{3-15}
L(t)dt= &&-iH_0(a,a^\dagger,t)dt+\sqrt{\gamma}(a
dB^\dagger_{\text{in}}(t)-dB_{\text{in}}(t)a^\dagger)- \frac{\gamma}
{2}a^\dagger adt\nonumber\\
&& -\frac{\gamma}{2}(a^\dagger a+aa^\dagger)dB_{\text{in}}^
\dagger(t)dB_{\text{in}}(t).
\end{eqnarray}
Initially the total wave-function $|\psi(0)\rangle\otimes|\{0\}
\rangle$ is the vacuum state $|\{0\}\rangle$ of the field. The total
wave function at time $t$ may be represented as the time-ordered
product over discretized time
\begin{equation}
\label{3-16}
\left[\prod_{t_i=0}^t(1+L(t_i)dt_i)\right]_+
|\psi(0)\rangle\otimes|\{0\}\rangle.
\end{equation}
Then we may use the fact $[dB_{\text{in}}(t),dB_{\text{in}}^\dagger
(t')]=0$ for $t\neq t'$ to bring all $dB(t_i)$ to
the right where they are annihilated when hitting $|\{0\}\rangle$. By
the same token we may replace $dB^\dagger_{\text{in}}(t_i)$
by
\begin{equation}
\label{3-17}
d\Xi(t_i)e^{-i\varphi}=dB^\dagger_{\text{in}}(t_i)+e^{-2i
\varphi}dB_{\text{in}}(t_i)
\end{equation}
which may be further replaced by $d\xi(t_i)e^{-i\varphi}$ in the
$\xi$-representation in which all $d\Xi(t)$ are simultaneously
diagonal. Thus $L(t)$ in eq.~(\ref{3-16}) is reduced to
\begin{equation}
\label{3-18}
L(t)dt=-iH_0(a,a^\dagger,t)dt-\frac{\gamma}{2}a^\dagger adt+
\sqrt{\gamma}ae^{-i\varphi}d\xi(t).
\end{equation}
At this stage there are no operators left in eq.~(\ref{3-16}) which
act on $|\{0\}\rangle$, which may therefore be {\it divided out}. The
essential point here is that the component of the wave-function
representing the environment (i.e. the vacuum) is {\it not}
eliminated by taking a trace, but rather by division. Therefore no
information is lost in this step.

For a single infinitesimal time-step the expression (\ref{3-16}) now
reduces to a linear stochastic wave-equation
\begin{equation}
\label{3-19}
d|\psi_\xi(t)\rangle=
\left[-iH_0(a,a^\dagger,t)dt-\frac{\gamma}{2}a^\dagger adt
  +\sqrt{\gamma}ae^{-i\varphi}d\xi(t)\right]|\psi_\xi(t)\rangle.
\end{equation}
In order to recover the {\it full} wave-equation, including the
vacuum which was eliminated , both sides of eq.~(\ref{3-19}) have to
be multiplied by $\varphi_0(\{\xi\})$, the vacuum $|\{0\}\rangle$ in
the $\xi$-representation. This functional is Gaussian and its
absolute square
\begin{equation}
\label{3-20}
|\varphi_0(\{\xi\})|^2d\{\xi\}=d\mu^W(\{\xi\})
\end{equation}
gives the classical Wiener measure of the white noise process
$\xi(s)$ with $\xi(t)=\int_0^td\xi(s)$, $(d\xi(s))^2=ds$. We shall
denote by $d\mu_t^W(\{\xi\})$ the Wiener measure of $\xi(s)$ for
$0<s\le t$. It should be noted that $|\psi_\xi(t)\rangle$ is not
normalized. However, the statistical operator defined by
$\rho(t)=\int d\mu^W_t(\{\xi\})\,|\psi_\xi(t)\rangle\langle\psi_\xi
(t)|$ is correctly normalized. If one wishes the usual master
equation for $\rho(t)$ can be derived from eq.~(\ref{3-19}), but
this is of no concern here.

{}From the unitarity of the total time-evolution with $H_{\text{int}}$
it follows that all expectation values of oscillator variables $Y$
are given by
\begin{eqnarray}
\label{3-21}
\langle Y(t)\rangle &=& \int|\varphi_0(\{\xi\})|^2 d\{\xi\}\langle
\psi_\xi(t)|Y|\psi_\xi(t)\rangle\nonumber\\
 &=& \int d\mu_t^W(\{\xi\})\langle\psi_\xi(t)|Y|\psi_\xi(t)\rangle.
\end{eqnarray}
More generally all time-ordered multi-time correlation functions of
not necessarily hermitian oscillator variables $Y_1$, $Y_2$, $Y_3$
etc.~are given by
\begin{equation}
\label{3-22}
\begin{array}{l}
\langle Y_1^\dagger(t_1')Y_2^\dagger(t_2')\dots Y_n^\dagger(t_n')
Y_m(t_m)\dots Y_2(t_2)Y_1(t_1)\rangle\\
\hspace{1cm}=\int d\mu^W(\{\xi\})\,\langle\psi_\xi(t'_1)|Y_1^\dagger
K_\xi^\dagger(t'_2-t'_1)Y_2^\dagger\dots K_\xi^\dagger(t'_n-t_{n-1}')
Y_n^\dagger L_\xi(t'_n-t_m)\times\\
\hspace{5.25cm}\times Y_mK_\xi(t_m-t_{m-1})\dots Y_2K_\xi(t_2-t_1)Y_1
|\psi_\xi(t_1)\rangle\end{array}
\end{equation}
where $t'_n>t'_{n-1}>\dots>t'_1$ and $t_m>t_{m-1}>\dots>t_1$. Here
$K_\xi(t)$ is the (non-unitary) time-evolution generated by
eq.~(\ref{3-19}) which may be written as
\begin{equation}
\label{3-23}
K_\xi(t)=\left[\exp\left\{-i\int_0^t\left(H_0-i\frac{\gamma}{2}
a^\dagger a-i\frac{\gamma}{2}a^2e^{-2i\varphi}\right)\,ds+
\sqrt{\gamma}ae^{-i\varphi}\int_0^td\xi(s)\right\}\right]_+.
\end{equation}
$K^\dagger_\xi(t)$ is the adjoint of $K_\xi(t)$. The kernel $L(t)$ in
the middle of eq.~(\ref{3-22}) is
\begin{equation}
\label{3-24}
L_\xi(t)=\left\{\begin{array}{ccl} K_\xi(t) & \text{if} & t>0\\
K_\xi^\dagger(t) & \text{if} & t<0\\
1 & \text{if} & t=0.\end{array}\right.
\end{equation}
Thus, the full information about the reduced dynamics of the quantum
oscillator is contained in the linear stochastic wave-equation
(\ref{3-19}). Even though $K_\xi(t)$ is nonunitary, it is unitary
under the integral over $d\mu_t^W(\{\xi\})$,
\begin{equation}
\int d\mu_t^W(\{\xi\})\, K_\xi^\dagger(t)K_\xi(t)=1.
\end{equation}
This may be checked explicitely but also follows immediately from
the unitarity of the time-evolution of the total system.

The full information about the output process $d\theta(t)$,
regardless of the internal dynamics of the system, is also contained
in eq.~(\ref{3-19}). Indeed, the measure $d\mu_t(\{\theta\})$ defined
by putting $Y_i=1=Y_i^\dagger$ in eq.~(\ref{3-22})
\begin{equation}
\label{3-25}
d\mu_t(\{\theta\}) = d\mu_t^W(\{\theta\})\langle\psi_\theta(t)
|\psi_\theta(t)\rangle
\end{equation}
can be identified with the measure of the output process for
$0<s\le t$. We note that eq.~(\ref{3-25}) is just a variant of
eq.~(\ref{2-5}) of section 2. To see the relation with the output
process one computes the conditional measure
\begin{equation}
\label{3-26}
p(d\theta(t))=\frac{d\mu_{t+dt}(\{\theta\})}{d\mu_t(\{\theta\})}
\end{equation}
in terms of the conditional Wiener measure
\begin{equation}
\label{3-27}
p^W(d\theta(t)) =\frac{d\mu^W_{t+dt}(\{\theta\})}{d\mu^W_t(\{
\theta\})}
\end{equation}
of the input process. One obtains
\begin{equation}
\label{3-28}
p(d\theta(t))=p^W(d\theta(t))
\bigg(1+\frac{d\langle\psi_\theta(t)|\psi_\theta(t)\rangle}
{\langle\psi_\theta(t)|\psi_\theta(t)\rangle}\bigg).
\end{equation}
{}From eq.~(\ref{3-19}) we obtain
\begin{equation}
\label{3-29}
\frac{d\langle\psi_\theta(t)|\psi_\theta(t)\rangle}
{\langle\psi_\theta(t)|\psi_\theta(t)\rangle}=
\sqrt{\gamma}
\frac{\langle\psi_\theta(t)|ae^{-i\varphi}+
a^\dagger e^{i\varphi}|\psi_\theta(t)\rangle}
{\langle\psi_\theta(t)|\psi_\theta(t)\rangle}d\theta(t).
\end{equation}
Recalling that $p^W(d\theta(t))$ is just a Gaussian, formally
written as
\begin{equation}
\label{3-30}
p^W(d\theta(t))=\frac{d\theta}{\sqrt{2\pi dt}}\exp
\left(-\frac{1}{2}\frac{d\theta(t)^2}{dt}\right),
\end{equation}
we obtain equally formally, using the Ito calculus with
$(d\theta(t))^2=dt$
\begin{equation}
\label{3-31}
p(d\theta(t)) =
\frac{d\theta}{\sqrt{2\pi dt}}\exp
\left[-\frac{1}{2dt}(d\theta-2\sqrt{\gamma}\langle X(t)\rangle_\theta
dt)^2\right].
\end{equation}
Therefore
\begin{equation}
\label{3-32}
d\theta(t)=2\sqrt{\gamma}\langle X(t)\rangle_\theta dt+d\xi(t)
\end{equation}
where $d\xi(t)$ is the Wiener increment and where we defined
\begin{equation}
\label{3-33}
\langle X(t)\rangle_\theta=
\frac{\langle\psi_\theta(t)|X|\psi_\theta(t)\rangle}
{\langle\psi_\theta(t)|\psi_\theta(t)\rangle}.
\end{equation}
Eqs.~(\ref{3-32}) and (\ref{3-31}) may be compared with
eqs.~(\ref{3-12}) and (\ref{Y}), respectively. In the latter
equations the output process $d\theta(t)$ was conditioned on the
simultaneous eigenvalue $x(t)$, while in  eqs.~(\ref{3-31}),
(\ref{3-32}) it is conditioned, via $\langle X(t)\rangle_{\theta}$,
on the values of $d\theta(s)$ for preceding times $s<t$. Therefore,
because their probability distributions differ, the increments
defined by eq.~(\ref{3-12}) and (\ref{3-32}) are not the same.
$\langle X(t)\rangle_\theta$ is the quantum expectation value of
the $X$-quadrature component for a given realization of the output
process at times prior to $t$.

Let us evaluate the multi-time correlation functions $\langle
d\theta(t_n)\dots d\theta(t_1)\rangle$ of the output process implied
by eq.~(\ref{3-25}) and (\ref{3-32}). In the measure (\ref{3-25}) of
$\theta(t)$ we use the representation
\begin{equation}
\label{E3}
|\psi_\theta(t_n)\rangle=K_\theta(t_n)|\psi(0)\rangle.
\end{equation}
In (\ref{3-25}) the integration is performed over the Wiener measure.
In order to achieve this explicitely, for each time $t_i$ appearing
in the multi-time correlation functions the infinitesimal step $dt_i$
in the functional (\ref{E3}) of $d\theta(t_i)$ is represented
explicitely as $(1+L(t_i))dt_i$ like in eq.~(\ref{3-16}) with

$L(t_i)$ from eq.~(\ref{3-18}), making the dependence on
$d\theta(t_i)$ explicit. The Wiener integrals over all the
$d\theta(t_i)$ can then easily be performed with the result
\begin{equation}
\label{E4}\langle d\theta(t_n)\dots d\theta(t_1)\rangle=
(2\sqrt{\gamma})^ndt_n\dots dt_1\langle:X(t_n)\dots X(t_1):\rangle
\end{equation}
where $:\,:$ denotes normal and time-ordering. The result (\ref{E4})
holds for arbitrary correlation functions as long as $t_i\neq t_j$.
For all pairs $t_i$, $t_j$ with $t_i=t_j$ one has to take
$d\theta(t_i)d\theta(t_i)=dt_i$ summing over all possible
pairings and consider the correspondingly reduced correlation
function.

In summary, the linear stochastic wave-equation has the following
uses:
\begin{enumerate}
\item[(i)] Solving it with $d\xi(t)$ representing Wiener noise it
can be used to obtain all correlation functions of system
(i.e.~oscillator) variables via eqs.~(\ref{3-22}).
\item[(ii)] Solving it after replacing $d\xi(t)\rightarrow d\theta
(t)$ representing the yet unknown output noise (i.e.~taking
$d\theta(t)$ arbitrary) one obtains the conditional quantum
expectation (\ref{3-33}) of the $X$-quadrature as a
functional of the output noise from eq.~(\ref{3-33}). All correlation
functions of the measured $X$-quadrature are then given by the
correlation functions (\ref{E4}) of (\ref{3-32}).
\item[(iii)] Finally the solution of (i) gives the complete measure
of the output field when inserted in eq.~(\ref{3-25}).
\end{enumerate}

If one is interested in generating stochastic numerical samples of
$\langle X(t)\rangle_\theta$ and hence $d\theta(t)$ one first makes
the replacement $d\xi(t)\rightarrow d\theta(t)$ in the linear
stochastic wave-equation and then reexpresses the unknown output
process $d\theta(t)$ by the known Wiener process $d\xi(t)$ via
eq.~(\ref{3-32}). However, this step generates a nonlinearity. The
superposition principle is then lost in these numerical simulations.
For nonlinear stochastic wave equations and numerical examples see
\cite{4,15,21,22}.

\section{Exactly solvable examples}
We consider two examples where eq.~(\ref{3-19}) can be exactly
solved.
The first example is a harmonic oscillator periodically driven by
some external field with amplitude $F$, and frequency $\omega=
\omega_0+\delta$ in the rotating wave approximation. Thus
\begin{equation}
\label{4-1}
H_0(a,a^\dagger,t)=\frac{\gamma}{2i}
\left(F^*e^{i\delta t}a-Fe^{-i\delta t}a^\dagger\right).
\end{equation}
With a coherent initial state
\begin{equation}
\label{4-2}
|\psi_\xi(0)\rangle=|\alpha\rangle,
\end{equation}
where
\begin{equation}
\label{4-3}
 a|\alpha\rangle=\alpha|\alpha\rangle,
\end{equation}
the solution of eq.~(\ref{3-19}), after some calculation, is obtained
as
\begin{eqnarray}
\label{4-4}
|\psi_\xi(t,\alpha)\rangle & = & \exp\left[\frac{1}{2}\left(|\alpha
(t)|^2-|\alpha(0)|^2+\alpha^2(t)-\alpha^2(0)\right)+\gamma\int_0^t ds
\,\alpha(s)\mbox{Re}\left(Fe^{-i\delta s}\right)\right.\nonumber\\
& & \left.+\sqrt{\gamma}\int_0^t d\xi(s)\,\alpha(s)e^{-i\varphi}
\right]|\alpha(t)\rangle
\end{eqnarray}
with
\begin{equation}
\alpha(t)=\left(\alpha-\frac{F}{1-\frac{2i\delta}{\gamma}}\right)
e^{-\frac{\gamma}{2}t}+\frac{F}{1-\frac{2i\delta}{\gamma}}e^{-i
\delta t}.
\end{equation}
As any initial state can be expanded in coherent states, with
$f(\alpha)=\frac{1}{\pi}\langle\alpha|\psi_\xi(0)\rangle$,
\begin{equation}
\label{4-5}
|\psi_\xi(0)\rangle=\int d^2\alpha\,f(\alpha)|\alpha\rangle
\end{equation}
and due to the linearity of eq.~(\ref{3-19}), the general solution
is simply
\begin{equation}
\label{4-6}
|\psi_\xi(t)\rangle=\int d^2\alpha\,f(\alpha)|\psi_\xi(t,\alpha)
\rangle.
\end{equation}

For a single coherent state as initial condition the measured
quadrature
\begin{equation}
\label{4-7}
\langle X(t)\rangle_\xi=\frac{\langle\psi_\xi(t,\alpha)|X|\psi_\xi
(t,\alpha)\rangle}{\langle\psi_\xi(t,\alpha)|\psi_\xi(t,\alpha)
\rangle}
=\frac{1}{2}\left[\alpha(t)e^{-i\varphi}+\mbox{c.~c.~}\right]
\end{equation}
turns out to be independent of the measurement noise. This is no
longer true if a superposition of coherent states is present
initially.

Using the present method the decay of the initial superposition
$|\psi_\xi(0)\rangle=\mbox{const.~}(|\alpha\rangle+|-\alpha
\rangle)$ is studied in \cite{23}. See also \cite{24,25} where this
problem is studied using the nonlinear stochastic wave-equation.

Finally, we determine the measure of the output process. Using
eq.~(\ref{3-25}) with $|\psi_\theta(t)\rangle$ given by
eq.~(\ref{4-6}) where $\xi$ is replaced by $\theta$, one obtains a
rather complicated expression. Unless $f(\alpha)$ is an exponential
linear, or quadratic plus linear, in the real- and imaginary parts
of $\alpha$, the resulting measure will be non-Gaussian. However,
if $\langle X(t)\rangle_\theta$ is known as a functional of
$\theta$, as for the example (\ref{4-7}) where it is even
independent of $\theta$, one may use eq.~(\ref{3-31}) to obtain
\begin{equation}
\label{4-8}
d\mu_t(\{\theta\})=d\mu_t^W(\{\theta\})\exp\left(2\sqrt{\gamma}
\int_0^t\langle X(s)\rangle_\theta d\theta(s)-2\gamma\int_0^t
\langle X(s)\rangle_\theta^2 ds\right).
\end{equation}
For the example (\ref{4-7}) the result (\ref{4-8}) is just a
shifted Wiener measure.

As a second example we turn to subharmonic generation from the
vacuum, where
\begin{equation}
\label{4-9}
H_0(a,a^\dagger)=\frac{\kappa}{2i}(a^2-a^{\dagger 2})
\end{equation}
and
\begin{equation}
\label{4-10}
|\psi_\xi(0)\rangle=|0\rangle.
\end{equation}
The ansatz
\begin{equation}
\label{4-11}
|\psi_\xi(t)\rangle=
\exp\left(g(t)+\alpha(t)a^\dagger+\frac{1}{2}\beta(t)a^{\dagger 2}
\right)|0\rangle
\end{equation}
solves eq.~(\ref{3-19}), provided the differential equations
\begin{eqnarray}
\label{4-12}
d\beta(t) &=&
-\gamma\beta(t)dt+\left(\kappa-(\kappa+\gamma e^{-2i\varphi})
  \beta^2(t)\right)dt\nonumber\\
d\alpha(t)&=&
-\left(\frac{\gamma}{2}+(\kappa+\gamma e^{-2i\varphi})\beta(t)\right)
\alpha(t)dt+\sqrt{\gamma}\beta(t)e^{-i\varphi}d\xi(t)\nonumber\\
dg(t) &=& -\frac{\kappa}{2}\left(\alpha^2(t)+\beta(t)\right)
dt-\frac{\gamma}{2}\alpha^2(t)e^{-2i\varphi}dt+\sqrt{\gamma}
 \alpha(t)e^{-i\varphi}d\xi(t)
\end{eqnarray}
are satisfied with the initial conditions
$\alpha(0)=\beta(0)=g(0)=0$. The integration of eqs.~(\ref{4-12})
yields
\begin{eqnarray}
\label{4-13}
\beta(t) &=& \frac{1}{\kappa+\gamma e^{-2i\varphi}}
\Bigg(-\frac{\gamma}{2}+\sqrt{\frac{\gamma^2}{4}+\kappa^2
   +\kappa\gamma e^{-2i\varphi}}\tanh
     \sqrt{\frac{\gamma^2}{4}+\kappa^2+\kappa\gamma
e^{-2i\varphi}}(t-t_0)\Bigg)\nonumber\\
\alpha(t) &=&
\sqrt{\gamma}\int_0^te^{-\frac{\gamma}{2}(t-\tau)-(\kappa+\gamma
e^{-2i\varphi})\int_\tau^t
\beta(s)ds}\beta(\tau)e^{-i\varphi}\,d\xi(\tau)\nonumber\\
g(t) &=& -\frac{1}{2}\int_0^t
 d\tau\,\left(\kappa\beta(\tau)+(\kappa+\gamma e^{-2i\varphi})
   \alpha^2(\tau)\right)+\sqrt{\gamma}\int_0^t\alpha(\tau)
    e^{-i\varphi}\,d\xi(\tau).
\end{eqnarray}
The coefficient $\beta(t)$ describes squeezing. It contains the
constant of integration $t_0$, which is fixed by the initial
condition $\beta(0)=0$. The squeezing coefficient $\beta(t)$
is influenced by the measurement through $\gamma$, but unlike
the coefficient $\alpha(t)$ which determines the amplitude and
is completely generated from the output-noise, the squeezing
coefficient $\beta(t)$ is independent of the noise.

It is easy to calculate the prediction for the measured quadrature
$\langle X(t)\rangle_\xi$ as a functional of the {\it output} noise,
here still written as $\xi$, for simplicity. We obtain
\begin{equation}
\label{4-14}
\langle X(t)\rangle_\xi = \frac{\mbox{Re}\left[\left(\alpha(t)+
\beta(t)\alpha^*(t)\right)e^{-i\varphi}\right]}{1-|\beta(t)|^2}
\end{equation}
where $\alpha$ is a complex linear functional of $\xi$. In fact
$\langle X(t)\rangle_\xi$ is also a linear functional of $\xi$.
The measure for the output noise is given by eq.~(\ref{3-25}) and
may be written as
\begin{equation}
\label{4-15}
 d\mu_t(\{\xi\}) = d{\mu_t}^W(\{\xi\})e^{g(t)+g^*(t)+|\alpha(t)|^2}
 \langle\alpha(t)|e^{\frac{1}{2}\beta^*(t)a^2}
   e^{\frac{1}{2}\beta(t)a^{\dagger2}}|\alpha(t)\rangle.
\end{equation}
The expectation value with respect to coherent states can
be evaluated explicitely by the general form of the Baker Haussdorff
formula, but we will not go into the necessary algebra here.
In fact, like in the first example there is an easier way directly
based on eq.~(\ref{3-31}) and the formula (\ref{4-14}), directly
yielding the Gaussian measure (\ref{4-8}) with
$\langle X(t)\rangle_\theta$ given by (\ref{4-14}), after $\xi$
is replaced by $\theta$.

\section{Heterodyne measurements}
In heterodyne measurements the frequency $\omega_{lo}$ of the
local oscillator differs from that of the quantum oscillator,
$\Omega=\omega_0-\omega_{lo}\neq 0$. Therefore the phase $\varphi$
in eq.~(\ref{3-6}) must be replaced by $\varphi+\Omega t$. In
particular we now define
\begin{eqnarray}
\label{5-1}
d\Theta(t)&=& \bigg(
 dB_{\text{out}}(t)e^{-i\varphi-i\Omega t}+dB^\dagger_{\text{out}}(t)
 e^{i\varphi+i\Omega t}\bigg)\nonumber\\
d\Xi(t) &=& \bigg(
 dB_{\text{in}}(t)e^{-i\varphi-i\Omega t}+dB^\dagger_{\text{in}}(t)
e^{i\varphi+i\Omega t}\bigg)\nonumber\\
X(t)&=& \frac{1}{2}\bigg(ae^{-i\varphi-i\Omega t}
 +a^\dagger e^{i\varphi+i\Omega t}\bigg)
\end{eqnarray}
and eqs.~(\ref{3-7}), (\ref{3-12}) hold unchanged. The entire
derivation of the preceding section leading up to eq.~(\ref{3-19})
may then be taken over, and we obtain
\begin{equation}
\label{5-2}
d|\psi_\xi(t)\rangle=
\bigg[-iH_0(a,a^\dagger,t)dt-\frac{\gamma}{2}a^\dagger adt+\sqrt
{\gamma}ae^{-i\varphi-i\Omega t}d\xi(t)\bigg]|\psi_\xi(t)\rangle
\end{equation}
which differs from eq.~(\ref{3-19}) only by the replacement
\begin{equation}
\label{5-3}
d\xi(t)\rightarrow d\xi(t)e^{-i\Omega t}.
\end{equation}

The examples of section 4 can be extended to the case of
heterodyning. However, it is not enough to make the replacement
(\ref{5-3}) in the solutions, because the Ito-rule $(d\xi(t))^2=dt$
is not analytic in $d\xi(t)$.

The wave-function (\ref{4-4}) of the harmonic oscillator is extended
to the case of heterodyning by the replacement
$\varphi\to\varphi(s)=\varphi+\Omega s$. As the measured quadrature
is independent of the noise it changes only because of the changed
definition (\ref{5-1}), again leading to the replacement
$\varphi\to\varphi+\Omega t$ in eq.~(\ref{4-7}).

In the example of subharmonic generation, the generalization to the
case of heterodyning is less straightforward. In eqs.~(\ref{4-12})
it leads to the same replacements $\varphi\to\varphi+\Omega t$ as
before. As a consequence the equation for the squeezing parameter
$\beta(t)$ can no longer be integrated by a separation of
variables. However, while $\beta(t)$ is changed it still remains
unaffected by the measurement noise. Furthermore, the solutions
for $\alpha(t)$ and $g(t)$ in eq.~(\ref{4-13}) remain valid if
the change $\varphi\to\varphi+\Omega t$ is made there. In the case
where $|\Omega|$ is large compared to the characteristic inverse
time-scales of the system it is possible to average over many
cycles of $\Omega$ \cite{8,9}. Then
\begin{equation}
\label{5-4}
e^{-i\varphi}e^{-i\Omega t}d\xi=d\xi_1+id\xi_2
\end{equation}
is a complex white noise with real- and imaginary parts
$d\xi_1$, $d\xi_2$ satisfying
\begin{equation}
\label{5-5}
d\xi_1^2=d\xi_2^2=\frac{1}{2}dt,\quad d\xi_1d\xi_2=0.
\end{equation}
The dependence on $\varphi$ and $\Omega$ in eq.~(\ref{5-2}) then
disappears. The two examples of section 4 then simplify again: In
the oscillator wave-function (\ref{4-4}) the term
$\alpha^2(s)e^{-2i\varphi(s)}$ averages out and disappears,
and $d\xi(s)e^{-i\varphi(s)}=d\xi_1(s)+id\xi_2(s)$ is now complex
white noise. The measured quadrature (\ref{4-7}) is unaffected by
these simplifications and produces a signal at frequencies
$\pm\Omega$.

In the example of subharmonic generation all terms with
$e^{-2i\varphi(t)}$ in eq.~(\ref{4-12}) disappear by averaging
out and $e^{-i\varphi(s)}d\xi(s)$ becomes complex white noise.
The solution is then given by (\ref{4-13}) without the
$e^{-2i\varphi}$-terms and with complex noise. The phase
$\varphi(t)=\varphi+\Omega t$ then only remains in
$\langle X(t)\rangle_\xi$ in eq.~(\ref{4-14}) and again produces
a signal at $\pm\Omega$.

\section{Photon counting}
In this section we consider the case of photon counting by an
ideal photodetector placed behind the partially transmitting mirror
of a cavity.
The concept of input and output processes used in the preceding
section to describe homodyne and heterodyne measurements can be
generalized to this case. Let us introduce the input photon number
process $\Lambda_{\text{in}}(t)$ defined by (see e.~g.~[24-26])
\begin{equation}
 \Lambda_{\text{in}}(t)=\int_0^t b^{\dagger}_{\text{in}}(s)
 b_{\text{in}}(s)ds
\end{equation}
where formally
\begin{equation}
 b_{\text{in}}(s)=\frac{dB_{\text{in}}(s)}{ds}.
\end{equation}
The increment $d\Lambda_{\text{in}}(t)$ can be expressed in
terms of the white-noise creation and annihilation operators as
\begin{equation}
\label{6-3}
 d\Lambda_{\text{in}}(t)=\frac{dB^{\dagger}_{\text{in}}(t)
 dB_{\text{in}}(t)}{dt}.
\end{equation}
It satisfies the commutation relations
\begin{equation}
 [d\Lambda_{\text{in}}(t),d\Lambda_{\text{in}}
 (t')]=0,\qquad [d\Lambda_{\text{in}}(t),X(s)]=0\quad\text{ for }\
 t\ge s
\end{equation}
for any system variable $X$. We shall denote the eigenvalues of
$d\Lambda_{\text{in}}(t)$ by $d\lambda(t)$.

The output number process can be calculated using the Hamiltonian
$H_{\text{int}}$ in the interaction representation of
eq.~(\ref{3-3}) as a unitary transformation
\begin{equation}
\label{6-1}
 \Lambda_{\text{out}}(t)=U^{\dagger}_{\text{int}}(t)
 \Lambda_{\text{in}}(t)U_{\text{int}}(t)
\end{equation}
of the input number process, with $U_{\text{int}}$ defined in
eq.~(\ref{3-4}). The increment $d\Lambda_{\text{out}}$ of the
output process satisfies the quantum stochastic differential equation
[24-26]
\begin{equation}
\label{6-2}
 d\Lambda_{\text{out}}(t)=d\Lambda_{\text{in}}(t)+
 \sqrt{\gamma}(a(t)dB^{\dagger}_{\text{in}}(t)+dB_{\text{in}}(t)
 a^{\dagger}(t))+\gamma a^{\dagger}(t)a(t)dt
\end{equation}
where all system operators are in the Heisenberg picture. Note that
the output photon number process satisfies the same equal-time
commutation relations and multiplication rule as the input process
\cite{p1,p3}. Eq.~(\ref{6-2}) is the analogue of eq.~(\ref{3-7}).
The difference lies in the fact that the output number process does
not commute with the input number process for equal times as can be
seen from eq.~(\ref{6-2}) and the commutation relation
$[dB_{\text{in}}(t),dB^{\dagger}_{\text{in}}(t)]=dt$;
as a consequence the input and output number processes at equal time
can not be diagonalized simultaneously, i.~e.~it is not possible to
express the eigenvalues of the output process in terms of the
$d\lambda(t)$.

Let us consider now the time evolution of the state of the total
system. The generator $L(t)dt$ is unchanged and still given by
eq.~(\ref{3-15}). Up to now we have not specified the initial state
of the bath which is necessary to evaluate the term $dB^{\dagger}_
{\text{in}}(t)dB_{\text{in}}(t)$. Assuming the bath to be in a
vacuum state as in section 3 we immediately see from
eq.~(\ref{6-3}) that the input number process vanishes, which is
clear from a physical point of view: due to its absorbing nature a
photodetector cannot 'see' the vacuum fluctuations. The mean of the
output process is given by $\langle d\Lambda_{\text{out}}(t)
\rangle=\gamma\langle a^{\dagger}a\rangle (t)dt$ as can be seen by
taking the expectation value of eq.~(\ref{6-2}) in the vacuum state
of the bath. However, in this case it is not clear how to apply the
input-output formalism to derive the equation governing the time
evolution of the state of the system. A calculation of this
time evolution can still be performed in a manner similar to the
preceding section if we take the bath to be in a coherent state
$|\{\beta\}\rangle$ with
\begin{equation}
\label{vv}
 dB_{\text{in}}(t)|\{\beta\}\rangle=\sqrt{\gamma}\epsilon(t)
 dt|\{\beta\}\rangle
\end{equation}
where $\sqrt{\gamma}\epsilon(t)$ is the arbitrary complex amplitude
of the coherent state. At the end of the calculation we may take the
limit $\epsilon\rightarrow 0$. Physically, this assumption means that
not only the vacuum fluctuations but also a small classical field
enter the cavity of the quantum oscillator. According to
eq.~(\ref{vv}) the product $dB^{\dagger}_{\text{in}}(t)dB_{\text{in}}
(t)$ is of second order in $dt$ and therefore can be set equal to
zero in the following (this substitution can be justified rigorously
in terms of quantum stochastic integration; see e.~g.~\cite{p5}). We
then obtain the following multiplication table
\begin{eqnarray}
 dB_{\text{in}}(t)dB_{\text{in}}(t) & = & 0=
 dB^{\dagger}_{\text{in}}(t) dB^{\dagger}_{\text{in}}
 (t)\nonumber\\
 dB^{\dagger}_{\text{in}}(t)dB_{\text{in}}(t) & = &
 0 \nonumber\\
 dB_{\text{in}}(t)dB^{\dagger}_{\text{in}}(t) & = &
 dt \nonumber\\
 dB_{\text{in}}(t)dt & = & 0=dB^{\dagger}_{\text{in}}
 (t)dt \nonumber\\
 d\Lambda_{\text{in}}(t)dB_{\text{in}}(t) & = & 0
 \nonumber\\
 d\Lambda_{\text{in}}(t)d\Lambda_{\text{in}}(t)
 & = & d\Lambda_{\text{in}}(t)\nonumber\\
 d\Lambda_{\text{in}}(t)dt & = & 0.
\end{eqnarray}
For the eigenvalues of $d\Lambda_{\text{in}}(t)=d\lambda(t)$
the condition
\begin{equation}
 (d\lambda(t))^2=d\lambda(t)
\end{equation}
implies
\begin{equation}
 d\lambda(t)=\left\{ \begin{array}{c} 0 \\ 1 \end{array} \right. .
\end{equation}
The $d\lambda(t)$ at different times are statistically independent
due to the statistical independence of the underlying quantum Poisson
process at different times.

Proceeding as in section 3 we compute the wave function of the total
system at time $t$ as
\begin{equation}
\label{6-10}
 |\psi_{\text{tot}}(t)\rangle= \left[ \prod_{t_i=0}^t (1+L(t_i)
 dt_i)\right]_+ |\psi(0)\rangle\otimes|\{\beta\}\rangle
\end{equation}
where we assume that initially the state of the total system
factorizes. Now the operators $dB_{\text{in}}(t_i)$ can be
commuted to the right where they are replaced by $\sqrt{\gamma}
\epsilon(t_i)dt_i$ when acting on $|\{\beta\}\rangle$. Then the
Schr\"odinger equation for the total wave function is given by
\begin{equation}
\label{6-4}
 \left(d+iH_0(a,a^\dagger,t)dt+\frac{\gamma}{2}a^{\dagger}adt
 -\sqrt{\gamma}(adB^{\dagger}_{\text{in}}(t)-
 \sqrt{\gamma}\epsilon(t) a^{\dagger}dt)\right)|\psi(t)
 \rangle\otimes|\{\beta\}\rangle=0.
\end{equation}
Now we make the transformations
\begin{eqnarray}
 \sqrt{\gamma}a dB^{\dagger}_{\text{in}}|\psi(t)\rangle
 \otimes|\{\beta\}\rangle & = & \sqrt{\gamma}a
 \frac{dB^{\dagger}_{\text{in}}(t)dB_{\text{in}}(t)}
 {\sqrt{\gamma}\epsilon(t)dt}|\psi(t)\rangle\otimes|\{\beta\}\rangle
 \nonumber\\
 & = & \frac{1}{\epsilon(t)}a d\Lambda_{\text{in}}(t)|
 \psi(t)\rangle\otimes|\{\beta\}\rangle
\end{eqnarray}
where we have used eq.~(\ref{6-3}) in the last step. In the
representation where the input number process
$d\Lambda_{\text{in}}(t)$ can be replaced by the eigenvalue
$d\lambda(t)$ the Schr\"odinger equation (\ref{6-4}) is therefore
given by
\begin{equation}
\label{6-13}
 \left(d+iH_0(a,a^\dagger,t)dt+\frac{\gamma}{2}a^{\dagger}adt+
 \gamma\epsilon(t)a^{\dagger}dt-\frac{1}{\epsilon(t)}ad\lambda(t)
 \right)|\psi_{\lambda}(t)\rangle\langle\lambda|\{\beta\}\rangle=0
\end{equation}
so that the bath state can be divided out. In this way we obtain a
linear stochastic Schr\"odinger equation for the state of the system
alone
\begin{equation}
\label{6-7}
 d|\psi_{\lambda}(t)\rangle=\left\{-iH_0(a,a^\dagger,t)dt-
 \frac{\gamma}{2}a^{\dagger}adt-\gamma\epsilon(t)a^{\dagger}dt+
 \frac{1}{\epsilon(t)}ad\lambda(t)\right\}|\psi_{\lambda}(t)\rangle
\end{equation}
with the measure
\begin{equation}
 d\mu^P(\{\lambda\})=|\langle\lambda|\{\beta\}\rangle|^2.
\end{equation}
Here and in the following $|\lambda\rangle$ denotes the number
states of the bath satisfying
\begin{equation}
\label{Z}
d\Lambda_{\text{in}}(s)|\lambda\rangle=d\lambda(s)|\lambda
\rangle
\end{equation}
for all $s$.
Discretizing time $t\rightarrow t_i$ and using the fact that
$d\lambda(t_i)$ is 0 or 1 we obtain the formal expression
\begin{equation}
\label{6-19}
 d\mu^P_t(\{\lambda\})=\prod^{N-1}_i
  \left[(1-\gamma|\epsilon(t_i)|^2 dt_i)
 (1-d\lambda(t_i))+\gamma|\epsilon(t_i)|^2 dt_i d\lambda(t_i)\right].
\end{equation}
Thus $d\lambda(t)$ is an independent Poisson process with mean value
\begin{equation}
\label{6-6}
 \langle d\lambda(t)\rangle=\langle\{\beta\}|d\Lambda_{\text{in}}(t)
 |\{\beta\}\rangle = \gamma|\epsilon(t)|^2 dt.
\end{equation}
Using eq.~(\ref{6-7}) and the fact that $d\lambda(t)dt=0$ we can
write the wave function of the system at time $t+dt$ in the form
\begin{eqnarray}
\label{6-5}
 |\psi_{\lambda}(t+dt)\rangle & = & |\psi_{\lambda}(t)\rangle+
 d|\psi_{\lambda}(t)\rangle\nonumber\\
 & = & (1-d\lambda(t))\left(1-iH_0(a,a^\dagger,t)dt-\frac{\gamma}{2}
 a^{\dagger}adt-\gamma\epsilon(t)a^{\dagger}dt\right)|\psi_{\lambda}
 (t)\rangle\nonumber\\
 & & +d\lambda(t)\left(\frac{a}{\epsilon(t)}+1\right)|\psi_{\lambda}
 (t)\rangle.
\end{eqnarray}
This expression clearly shows how the wave function evolves during
the time interval $dt$ depending on the result of the photon
counting: if no photon is detected (with probability $1-\gamma|
\epsilon(t)|^2dt$ according to eq.~(\ref{6-19})), i.~e.~$d\lambda
(t)=0$, this time evolution, according to eq.~(\ref{6-5}), is
governed by the non-hermitian Hamiltonian $\tilde{H}=H_0-i\frac
{\gamma}{2}a^{\dagger}a-i\gamma\epsilon(t)a^{\dagger}$; if a photon
is detected (with probability $\gamma|\epsilon(t)|^2dt$), i.~e.~$d
\lambda(t)=1$, the wave function jumps according to the second term
in eq.~(\ref{6-5}). This time evolution of the wave function due to
the linear stochastic Schr\"odinger equation (\ref{6-7}) is analogous
to the nonlinear prescription introduced phenomenologically by
Carmichael \cite{15} to generate numerically what he has called
'quantum trajectories' for the wave function of the system
conditioned on the measured photon number (see also \cite{8}).
Applying eq.~(\ref{6-5}) repeatedly in time the formal solution of
the linear stochastic Schr\"odinger equation (\ref{6-7}) given by
eq.~(\ref{6-10}) can be written in the more elegant form, with
discretized time $t_i$, $t_0=0$, $t_N=t$
\begin{eqnarray}
\label{6-11}
 |\psi_{\lambda}(t)\rangle & = & \left[\prod_{i=0}^{N-1}\left\{(1-
 d\lambda(t_i))\left(1-iH_0(a,a^\dagger,t)dt-\frac{\gamma}{2}
 a^{\dagger}adt_i-\gamma\epsilon(t_i)a^{\dagger}dt_i\right)\right.
 \right.\nonumber\\
 & & \left.\left.+d\lambda(t_i)\left(\frac{a}{\epsilon(t_i)}+1
 \right)\right\}\right]_+|\psi(0)\rangle\nonumber\\
 & \equiv & K_\lambda(t,0)|\psi(0)\rangle
\end{eqnarray}
which will be useful for the calculation of correlation functions. It
is important to note that as a consequence of factoring out the state
of the bath in eq.~(\ref{6-13}) the time-evolution operator
$K_\lambda(t,0)$ is non-unitary. It is now straightforward to obtain
the analogue of eq.~(\ref{3-22}) for arbitrary multi-time correlation
functions with $d\mu^W$ replaced by $d\mu^P$.

Let us show now how the complete information on the output process is
contained in the solution of eq.~(\ref{6-7}). To this end we define a
jump process $\kappa(t)$, with $(d\kappa(t))^2=d\kappa(t)$, which is
distributed according to the measure
\begin{equation}
\label{6-8}
 d\mu_t(\{\kappa\})=d\mu^{P}_t(\{\kappa\})\langle\psi_{\kappa}(t)|
 \psi_\kappa(t)\rangle
\end{equation}
where $d\mu^{P}_t(\{\kappa\})$ is the Poisson measure (\ref{6-19}) of
the input process. The measure (\ref{6-8}) is normalized, i.~e.
\begin{eqnarray}
 \int d\mu_t(\{\kappa\}) & = & \int d\{\kappa\}\,|\langle\{\beta\}|
 \kappa\rangle|^2 \langle\psi_\kappa(t)|\psi_\kappa(t)\rangle
 =\int d\{\kappa\}\,\langle\psi_\kappa(t)|\otimes\langle\{\beta\}|
 \kappa\rangle\langle\kappa|\{\beta\}\rangle\otimes|\psi_\kappa(t)
 \rangle\nonumber\\
 & = & \int d\{\kappa\}\langle\psi_{\text{tot}}(t)|\kappa
 \rangle\langle\kappa|\psi_{\text{tot}}(t)\rangle
 =\langle\psi_{\text{tot}}(t)|\psi_{\text{tot}}(t)\rangle
 =1
\end{eqnarray}
because the total wave-function of system and environment is
normalized. Going through the same argument as in section 3,
eq.~(\ref{3-25})-(\ref{3-33}), we obtain for the conditional measure
of a single time step
\begin{equation}
 p(d\kappa(t))=p^P(d\kappa(t))\left(1+\frac{d\langle\psi_{\kappa}(t)
 |\psi_{\kappa}(t)\rangle}{\langle\psi_{\kappa}(t)|\psi_{\kappa}(t)
 \rangle}\right)
\end{equation}
where
\begin{equation}
\label{U}
p^P(d\kappa(t))=(1-\gamma|\epsilon|^2dt)(1-d\kappa(t))
+\gamma|\epsilon|^2dtd\kappa(t).
\end{equation}
The increment of the norm of $|\psi_{\kappa}(t)\rangle$ can be
calculated from eq.~(\ref{6-5}) and we obtain
\begin{eqnarray}
\label{6-15}
 p(d\kappa(t)=0) & = & (1-\gamma|\epsilon(t)|^2dt)\left(1-\gamma
 \langle\epsilon(t)a^{\dagger}+\epsilon^{\star}(t)a\rangle_\kappa
 dt-\gamma\langle a^{\dagger}a\rangle_\kappa dt\right)\nonumber\\
 & = & 1-\gamma|\epsilon(t)|^2dt-\gamma\langle\epsilon(t)
 a^{\dagger}+\epsilon^{\star}(t)a\rangle_\kappa dt-\gamma\langle
 a^{\dagger}a\rangle_\kappa dt\nonumber\\
 p(d\kappa(t)=1) & = & \gamma|\epsilon(t)|^2dt\left(1+\frac{\langle
 a\rangle_\kappa}{\epsilon(t)}+\frac{\langle a^{\dagger}\rangle_
 \kappa}{\epsilon^{\star}(t)}+\frac{\langle a^{\dagger}a\rangle_
 \kappa}{|\epsilon(t)|^2}-\gamma\langle\epsilon(t)a^{\dagger}+
 \epsilon^{\star}(t)a\rangle_\kappa dt-\gamma\langle a^{\dagger}a
 \rangle_\kappa dt\right)\nonumber\\
 & = & \gamma|\epsilon(t)|^2dt+\gamma\langle\epsilon(t)a^{\dagger}+
 \epsilon^{\star}(t)a\rangle_\kappa dt+\gamma\langle a^{\dagger}a
 \rangle_\kappa dt
\end{eqnarray}
where we defined for an arbitrary operator $\Omega$
\begin{equation}
 \langle\Omega(t)\rangle_\kappa=\frac{\langle\psi_\kappa(t)|\Omega|
 \psi_\kappa(t)\rangle}{\langle\psi_\kappa(t)|\psi_\kappa(t)\rangle}.
\end{equation}
The expectation value of $d\kappa(t)$ according to this normalized
probability distribution is simply given by
\begin{equation}
\label{6-9}
 \langle d\kappa(t)\rangle=p(d\kappa(t)=1).
\end{equation}
Comparing eq.~(\ref{6-15}) with the expectation value of
eq.~(\ref{6-2}) in the coherent state of the bath we see that the
expectation value of the process $d\kappa(t)$ introduced by
eq.~(\ref{6-8}) is equal to that of the output process $d\Lambda_
{\text{out}}(t)$.

We can go even further and show the equality of all correlation
functions of $d\kappa(t)$ and the corresponding correlation
functions of $d\Lambda_{\text{out}}(t)$, which means that
$d\kappa(t)$ is stochastically equivalent to the output process.
The calculation of correlation functions of $d\kappa(t)$ is greatly
simplified by the use of the formal solution (\ref{6-11}) of the
linear stochastic Schr\"odinger equation (\ref{6-7}) (see also
section 3). Let us consider first the two-time correlation function
\begin{equation}
\label{V}
\int d\mu_{t+\tau}(\{\kappa\})(d\kappa(t+\tau)d\kappa(t))=
\langle d\kappa(t+\tau)d\kappa(t)\rangle
\end{equation}
with $\tau>0$. According to eq.~(\ref{6-8})
we can reexpress eq.~(\ref{V}) in terms of the two-time
correlation function of the input process as
\begin{equation}
\label{6-12}
 \langle d\kappa(t+\tau)d\kappa(t)\rangle=\int  d\mu_{t+\tau}^P
 (\{\lambda\})\,d\lambda(t+\tau)d\lambda(t)\langle\psi_{\lambda}
 (t+\tau+d\tau)|\psi_\lambda(t+\tau+d\tau)\rangle
\end{equation}
where $|\psi_\lambda(t+\tau+d\tau)\rangle$ is given by
eq.~(\ref{6-11}). Now we assume that the wave function at time $t$ is
given and not conditioned on the input process, so that $|\psi(t)
\rangle$ is simply the initial state of the system. The components of
$|\psi_\lambda(t+\tau+d\tau)\rangle$ with $d\lambda(t)=0$ or
$d\lambda(t+\tau)=0$ cannot contribute in the integral (\ref{6-12}).
Therefore we can rewrite eq.~(\ref{6-12}) as
\begin{eqnarray}
\label{6-14}
 \langle d\kappa(t+\tau)d\kappa(t)\rangle & = & \int d\mu_{t+\tau}^P
 (\{\lambda\}|d\lambda(t)=d\lambda(t+\tau)=1)\,\langle\psi_\lambda(t
 +\tau+d\tau)|\psi_\lambda(t+\tau+d\tau)\rangle\times\nonumber\\
 & & \times p^P(d\lambda(t)=1)p^P(d\lambda(t+\tau)=1)
\end{eqnarray}
with the Poisson measure conditioned on $d\lambda(t)=d\lambda(t+\tau)
=1$ and
\begin{equation}
 |\psi_\lambda(t+\tau+d\tau)\rangle=\left(\frac{a}{\epsilon(t+\tau)}
 +1\right)K_\lambda(t+\tau,t+dt)\left(\frac{a}{\epsilon(t)}+1\right)
 |\psi(t)\rangle.
\end{equation}
The non-unitary evolution operator $K_\lambda(t+\tau,t+dt)$ is
defined in eq.~(\ref{6-11}) and we have used again the fact that
$d\lambda(t)=d\lambda(t+\tau)=1$. Since the probability $p^P(d\lambda
(t)=1)$ is independent of the state $|\psi_\lambda(t)\rangle$ of the
system we obtain
\begin{eqnarray}
 \label{ab}
 \langle d\kappa(t+\tau)d\kappa(t)\rangle & = & \gamma^2|\epsilon(t)
 |^2|\epsilon(t+\tau)|^2 dt^2\int d\mu_{t+\tau}^P(\{\lambda\})\,
 \langle\psi(t)|\left(\frac{a^\dagger}{\epsilon^\star(t)}+1\right)
 K_\lambda^\dagger(t+\tau,t+dt)\times\nonumber\\
 & & \times\left(\frac{a^\dagger}{\epsilon^\star(t+\tau)}+1\right)
 \left(\frac{a}{\epsilon(t+\tau)}+1\right)
 K_\lambda(t+\tau,t+dt)
 \left(\frac{a}{\epsilon(t)}+1\right)|\psi(t)\rangle\nonumber\\
 & = & \gamma^2 dt^2\int d\mu_{t+\tau}^P(\{\lambda\})\,\langle
 \psi(t)|(a^\dagger+\epsilon^\star(t))K^\dagger_\lambda(t+\tau,
 t+dt)(a^\dagger+\epsilon^\star(t+\tau))\times\nonumber\\
 && \times(a+\epsilon(t+\tau))K_\lambda(t+\tau,t+dt)|\psi(t)\rangle.
\end{eqnarray}
Due to the non-unitarity of $K_\lambda(t+\tau,t+dt)$ it is not
possible, in general, to insert $K_\lambda K_\lambda^\dagger$
between factors at will. However, because the {\it total} time
evolution of system and reservoir is unitary, we still have
unitarity under the integral
\begin{equation}
 \int d\mu_{t+\tau}^P(\{\lambda\})\,K^\dagger_\lambda(t+\tau,t+dt)
  K_\lambda(t+\tau,t+dt)=1
\end{equation}
as may be checked by using eq.~(\ref{6-11}). Therefore {\it under
the integral} over $d\mu^P(\{\lambda\})$ the stochastic
time-evolution operator $K_\lambda(t+\tau,t+dt)$ may be treated as
unitary, and in this sense $K_\lambda (t)$ can be used to define a
stochastic Heisenberg picture,
\begin{equation}
\label{A}
\Omega_\lambda(t)=K^\dagger_{\lambda}(t,0+dt)\Omega K_\lambda(t,
 0+dt).
\end{equation}
Using this we may now rewrite eq.~(\ref{ab}) as
\begin{equation}
 \langle d\kappa(t+\tau)d\kappa(t)\rangle=\gamma^2 dt^2\int
 d\mu_{t+\tau}^P(\{\lambda\})\,\langle\psi|\tilde{a}^\dagger_\lambda
 (t)\tilde{a}^\dagger_\lambda(t+\tau)\tilde{a}_\lambda(t+\tau)\tilde
 {a}_\lambda(t)|\psi\rangle
\end{equation}
where we used $\tilde{a}(t)=a(t)+\epsilon(t)$. The right-hand side
is the standard form for the degree of second order coherence
$g^{(2)}(t,t+\tau)$ to be found in textbooks on quantum optics
(see e.~g.~\cite{p8}).

The above procedure can be generalized in a straightforward manner
for the calculation of correlation functions of higher order. The
general $n$-time correlation function for unequal times $t_i\ne t_j$
for all $i,j$ is thereby obtained as
\begin{equation}
\label{6-18}
 \langle d\kappa(t_n)\ldots d\kappa(t_1)\rangle=\gamma^n dt^n\int
 d\mu_{t_n}^P(\{\lambda\})\,\langle\psi|\tilde{a}^\dagger_\lambda
 (t_1)\ldots\tilde{a}^\dagger_\lambda(t_n)\tilde{a}_\lambda(t_n)
 \ldots \tilde{a}_\lambda(t_1)|\psi\rangle
\end{equation}
where we have assumed that $t_n> t_{n-1}>\ldots t_1$.
Eq.~(\ref{6-18}) is equal to the degree of $n$-th order coherence.

To complete the comparison of the process $d\kappa(t)$ with the
output process $d\Lambda_{\text{out}}(t)$ we compute the
general normal ordered $n$-time correlation function of the output
process
\begin{equation}
 \langle :d\Lambda_{\text{out}}(t_n)\ldots d\Lambda_{\text{out}}
 (t_1):\rangle =\langle\psi(t_1)|\otimes\langle\{\beta
 \}|:d\Lambda_{\text{out}}(t_n)\ldots d\Lambda_{\text{out}}(t_1):|
 \{\beta\}\rangle\otimes |\psi(t_1)\rangle
\end{equation}
where $:\,:$ denotes normal and time-ordering. Note that the
expectation value is taken in the state of the total system because
the output process $d\Lambda_{\text{out}}(t)$ acts in the
Hilbert space of the total system. Expressing $d\Lambda_{\text{out}}
(t)$ in terms of the input field by eq.~(\ref{6-2}) and
using the statistical independence of $d\Lambda_{\text{in}}(t)
$, $dB_{\text{in}}(t)$ and $dB^\dagger_{\text{in}}(t)$
at different times we finally obtain
\begin{equation}
\label{6-17}
 \langle :d\Lambda_{\text{out}}(t_n)\ldots d\Lambda_{\text{out}}
 (t_1):\rangle=\gamma^n dt^n \langle \tilde{a}^\dagger
 (t_1)\ldots \tilde{a}^\dagger(t_n)\tilde{a}(t_n)\ldots\tilde{a}
 (t_1)\rangle
\end{equation}
where the expectation value on the right side of eq.~(\ref{6-17})
is again taken in the state of the total system. Now comparing
eq.~(\ref{6-17}) with eq.~(\ref{6-18}) we see that both expressions
are equal and $d\kappa(t)$ is indeed statistically equivalent to the
output process.

Let us now make contact to the nonlinear stochastic Schr\"odinger
equation used by others. This is done by replacing the input process
$d\lambda(t)$ in eq.~(\ref{6-7}) by the output process $d\kappa(t)$,
i.~e.
\begin{equation}
\label{bb}
 \frac{d\lambda(t)}{\sqrt{\langle d\lambda(t)\rangle}}\rightarrow
 \frac{d\kappa(t)}{\sqrt{(\gamma(|\epsilon(t)|^2+\langle\epsilon(t)
 a^{\dagger}+\epsilon^{\star}(t)a\rangle_\kappa+\langle a^{\dagger}
 a\rangle_\kappa))dt}}\equiv\frac{d\kappa(t)}{\sqrt{P(t)dt}}
\end{equation}
where $P(t)dt=\langle d\kappa(t)\rangle$. This replacement is
phenomenological in distinction to the microscopic derivation of
eq.~(\ref{6-7}) we have given above. It corresponds to the transition
from the linear equation (\ref{2-2}) to the nonlinear equation
(\ref{1-3}) by taking $\tilde{p}_{\tilde{N}}$ as equal to the ouput
probability $|\langle\psi_{\tilde{N}}|\psi\rangle|^2$. In this way
we obtain a {\it nonlinear} stochastic Schr\"odinger equation for
the non-normalized wave function $|\psi_\kappa(t)\rangle$
\begin{equation}
\label{6-21}
 d|\psi_\kappa(t)\rangle=\left\{-iH_0(a,a^\dagger,t)
 dt-\frac{\gamma}{2}a^{\dagger}adt-\gamma\epsilon(t)
 a^{\dagger}dt+\sqrt{\gamma}\frac{a}
 {\sqrt{P(t)}}d\kappa(t)\right\}|\psi_\kappa(t)\rangle.
\end{equation}
The nonlinearity arises because $P(t)$ depends on the wave-function
due to the appearance of expectation values in its definition
(\ref{bb}). The superposition principle is thereby destroyed.
Performing the limit $\epsilon(t)\rightarrow 0$, i.~e.~setting the
coherent driving field equal to zero this stochastic Schr\"odinger
equation takes on the form
\begin{equation}
\label{6-22}
 d|\psi_\kappa(t)\rangle=\left\{-iH_0(a,a^\dagger,t)dt-\frac{\gamma}
 {2}a^{\dagger}adt+\frac{a}{\sqrt{\langle a^{\dagger}a\rangle_
 \kappa}}d\kappa(t)\right\}|\psi_\kappa(t)\rangle
\end{equation}
and for the normalized $|\phi_\kappa(t)\rangle=|\psi_\kappa(t)\rangle
/\langle\psi_\kappa(t)|\psi_\kappa(t)\rangle^{1/2}$
\begin{equation}
\label{6-23}
 d|\phi_\kappa(t)\rangle=\left\{-iH_0(a,a^\dagger,t)
 dt-\frac{\gamma}{2}(a^{\dagger}a-\langle a^{\dagger}a\rangle)dt+
 \left(\frac{a}{\sqrt{\langle a^{\dagger}a\rangle_\kappa}}-1\right)
 d\kappa(t)\right\}|\phi_\kappa(t)\rangle
\end{equation}
which is identical to the nonlinear stochastic Schr\"odinger equation
introduced by Carmichael \cite{15} (see also \cite{9}). Note that the
classical output process cannot be reexpressed in terms of the
classical input process as was already stated in the beginning of
this section. It should also be noted that the limit $\epsilon
\rightarrow 0$ obviously cannot be taken in the linear equation
(\ref{6-7}). However, even in eq.~(\ref{6-21}) this limit cannot be
taken in a strict form, because the resulting eqs.~(\ref{6-22}),
(\ref{6-23}) are ill defined for the vacuum state, where $\langle
a^\dagger a\rangle_\kappa=0$.

Finally, for illustration we apply the linear stochastic
Schr\"odinger equation (\ref{6-7}) to the periodically driven
harmonic oscillator already studied in section 4, i.~e.~we solve
the linear stochastic Schr\"odinger equation
\begin{equation}
\label{6-24}
 d|\psi_\lambda(t)\rangle=\left\{ -\frac{\gamma}{2}\left(F^\star
 e^{i\delta t}a-Fe^{-i\delta t}a^\dagger\right)dt-\frac{\gamma}{2}
 a^\dagger adt-\gamma\epsilon(t)a^\dagger dt+\frac{a}{\epsilon(t)}
 d\lambda(t)\right\}|\psi_\lambda(t)\rangle
\end{equation}
where initially $|\psi(0)\rangle=|\alpha_0\rangle$. Inserting the
ansatz $|\psi_\lambda(t)\rangle=g(t)|\alpha(t)\rangle$ in
eq.~(\ref{6-24}) gives the following two differential equations
\begin{eqnarray}
 d\alpha(t) & = & -\left( \frac{\gamma}{2}\alpha(t)+\gamma\epsilon(t)
 -\frac{\gamma}{2}Fe^{-i\delta t}\right)dt\nonumber\\
 dg(t) & = & \left[-\frac{\gamma}{2}\left\{ |\alpha(t)|^2+i\mbox{Im}
 \left(F^\star\alpha(t)e^{i\delta t}\right)+2\mbox{Re}(\alpha^\star
 (t)\epsilon(t))\right\}dt+\frac{\alpha(t)}{\epsilon(t)}d\lambda(t)
 \right]g(t)
\end{eqnarray}
with initial conditions $g(0)=1$, $\alpha(0)=\alpha_0$. The
integration of the differential equation for $\alpha(t)$ yields
\begin{equation}
 \alpha(t)=\left(\alpha_0-\frac{F}{1-\frac{2i\delta}{\gamma}}\right)
 e^{-\frac{\gamma}{2}t}+\frac{F}{1-\frac{2i\delta}{\gamma}}
 e^{-i\delta t}-\gamma e^{-\frac{\gamma}{2}t}\int_0^t ds\,\epsilon(s)
 e^{\frac{\gamma}{2}s}
\end{equation}
while the solution for $g(t)$ can be written in the manner of
eq.~(\ref{6-11}) as
\begin{eqnarray}
 g(t) & = & \prod_{i=0}^{N-1}\left[(1-d\lambda(t_i))\left(-\frac
 {\gamma}{2}\left\{ |\alpha(t_i)|^2 +i\mbox{Im}\left(F^\star
 \alpha(t_i)e^{i\delta t_i}\right)+2\mbox{Re}(\alpha^\star(t_i)
 \epsilon(t_i))\right\}dt\right)\right.\nonumber\\
 & & \left. +d\lambda(t_i)\frac{\alpha(t_i)}{\epsilon(t_i)}\right]
\end{eqnarray}
with $t_0=0$, $t_N=t$.

\section{Conclusion}
Traditionally the time evolution of quantum systems comes in two
forms, one is linear, unitary, and deterministic, and describes
systems which are completely isolated, even from any external
measuring apparatus; the other is nonlinear, non-unitary and random
and describes the results of measurements. In the present paper we
have discussed how the nonlinearity in the description of
measurements can be avoided, first for the usual schematic and
idealized general description of quantum measurements, and then more
concretely for the well-known quantum optical measurement schemes of
homodyne- and heterodyne measurements and photon-counting. In each of
these concrete examples we have given microscopic derivations of the
linear wave-equations, which govern the measured systems. In the
limit of vanishing coupling with the measurement devices $(\gamma
\to 0)$ they reduce to the Schr\"odinger equation for closed systems.
The existence of these equations should dispel the old believe that
the two forms of quantum dynamics mentioned above are incompatible.
Rather they appear here as limiting forms of the microscopically
founded description we have given. This is only possible because a
{\it linear} description of non-isolated measured quantum systems
exists, which preserves the validity of the superposition principle
also for this class of systems. Besides some practical advantages in
certain special cases this is the fundamental reason why we have put
all emphasis on the linear form of the wave-equations and mentioned
their equivalent nonlinear counterparts only in passing. In this
linear description the effects of the environment or the measurement
apparatus resides only in the non-unitary dissipative terms and the
non-unitary stochastic terms of the wave-equation. These alone are
sufficient to produce the familiar disappearance of interference
terms as the coupling of the system to the external world is
increased, a nonlinearity of any form is not required to produce
this effect.

For the three measurement schemes which we have discussed in detail
we have shown that the solutions of the linear wave-equations yield
\begin{itemize}
\item expectation values and correlation functions of all system
variables from formulas like eq.~(\ref{3-22}), where the unnormalized
stochastic expectation value taken with the stochastic wave-function
is averaged over the Wiener noise, in the case of homodyne or
heterodyne measurements, or over the Poisson noise in the case of
photon counting;
\item stochastic realizations according to their correct measure of
the measured output field $\Theta(t)$ given by eqs.~(\ref{3-6}) or
(\ref{5-1}), or $\Lambda_{\text{out}}(t)$ of eqs.~(\ref{6-1}),
(\ref{6-2}) in the case of photon-counting, via the normalized
stochastic expectation values $\langle X(t)\rangle_\theta$ or
$\langle\tilde{a}^\dagger(t)\tilde{a}(t)\rangle_\kappa$ in
eqs.~(\ref{3-31}) or (\ref{6-15}), respectively; this result is
particularly important because it allows to evaluate {\it all}
correlation-functions, power spectra, etc.~of the measured quantities
from a single, sufficiently long time series via eqs.~(\ref{E4}) or
(\ref{6-18}), respectively.
\item the complete measure of the output field of the measurement
via eq.~(\ref{3-25}), or eq.~(\ref{6-8}), respectively.
\end{itemize}

The {\it practical} advantage of the linear description we have
illustrated by giving exact solutions for a number of examples. Of
course many more examples for exact solutions can and undoubted will
now be given: all examples of nonlinear quantum optics which allow
for a linearization (e.~g.~linear parametric amplifiers, Raman
amplifiers, nonlinear quantum oscillators like lasers if operating
far above threshold) fall in this class. However, the linear
stochastic wave-equations present crucial advantages even if
intrinsicly nonlinear physical processes are considered whose quantum
effects under measurements or interactions with reservoirs are of
interest --- e.~g.~the important class of nonlinear mesoscopic
systems exhibiting at the same time quantum interference effects like
dynamical localization or Ericson-type fluctuations and traces of
classical chaos. Obviously, if stochastic wave-equations are
chosen to describe such systems it is only their linear version which
will permit to give a transparent description of such quantum
interference effects. For instance path-integral solutions of the
linear wave-equations exist and semi-classical approximations can be
based on them.

So far the theory has only been developed within the
Markovian framework. In fact, the reliance on the Markovian limit is
heavy and one may suspect that this limit is really essential in
order to achieve the description of the subsystem by a stochastic
wave-function. Although this may be possible, it has not been shown,
and non-Markovian generalizations of the formalism we have presented
remain a challenge.

\acknowledgements
One of us (R.~G.) wishes to acknowledge useful remarks by Crispin
Gardiner, Mathew Collet and Howard Carmichael on the occasion of the
symposium 'Quantum Optics VI' in Rotorua, New Zealand, where parts of
this work were presented. This work was supported by the Deutsche
Forschungsgemeinschaft through the Sonderforschungsbereich 237
"Unordnung und gro{\ss}e Fluktuationen".



\newpage
 \begin {thebibliography}{99}
  \bibitem {1} N.~Gisin, {\it Phys.~Rev.~Lett. \bf 52}, 1657
   (1984); {\it Helv.~Phys.~Acta \bf 62}, 363 (1989).
  \bibitem {2} P.~Pearle, {\it Phys.~Rev.~D \bf 13}, 857
   (1976).
  \bibitem {3} L.~Di\'osi, {\it Phys.~Lett. A \bf129}, 419
    (1988).
  \bibitem {4} N.~Gisin, I.~C.~Percival, {\it J.~Phys.~A} {\bf 25},
        5677 (1992); {\it Phys.~Lett.~A} {\bf 167}, 315 (1992).
  \bibitem {5} G.~C.~Ghirardi, P.~Pearle, A.~Rimini,
   {\it Phys.~Rev.~A \bf 42}, 78 (1990).
  \bibitem {8} H.~M.~Wiseman, G.~J.~Milburn, {\it Phys.~Rev.~A}
   {\bf 47}, 642 (1993).
  \bibitem {9} H.~M.~Wiseman, G.~J.~Milburn, {\it Phys.~Rev.~A}
   {\bf 47}, 1652 (1993).
  \bibitem {7} V.~P.~Belavkin, {\it J.~Math.~Phys.~} {\bf 31}, 2930
   (1990).
  \bibitem {11} V.~P.~Belavkin, P.~Staszewski, {\it Phys.~Rev.~A}
   {\bf 45}, 1347 (1992).
  \bibitem {6} V.~P.~Belavkin, {\it J.~Phys.~A} {\bf 22}, L1109
   (1989).
  \bibitem {10} V.~P.~Belavkin, {\it Phys.~Lett.~A} {\bf 140}, 355
   (1989).
  \bibitem {14} A.~Barchielli, preprints.
  \bibitem {13} A.~Barchielli, V.~P.~Belavkin, {\it J.~Phys.~A}
   {\bf 24}, 1495 (1991).
  \bibitem {Z} C.~W.~Gardiner, A.~S.~Parkins, P.~Zoller,
           {\it Phys.~Rev.~A \bf 46}, 4363 (1992);\\
           R.~Dum, A.~S.~Parkins, P.~Zoller, C.~W.~Gardiner,
           {\it Phys.~Rev.~A \bf 46}, 4382 (1992).
  \bibitem {15} H.~J.~Carmichael, {\it An Open Systems Approach to
   Quantum Optics}, Lecture Notes in Physics m18, Berlin, Springer
   1993.
  \bibitem {16} J.~von Neumann, {\it Mathematische Grundlagen der
   Quantenmechanik}, Berlin, Springer 1932.
  \bibitem {17} E.~P.~Wigner, {Am.~J.~Phys.} {\bf 31}, 6 (1963).
  \bibitem {18} M.~J.~Collet, C.~W.~Gardiner, {\it Phys.~Rev.~A}
   {\bf 30}, 1386 (1984);\\
   C.~W.~Gardiner, M.~J.~Collet, {\it Phys.~Rev.~A} {\bf 31}, 3761
   (1985).
  \bibitem {19} C.~W.~Gardiner, {\it Quantum Noise}, Berlin, Springer
   1991.
  \bibitem {21} P.~Goetsch, R.~Graham, {\it Ann. Physik \bf 2}, 706
   (1993).
  \bibitem {22} B.~M.~Garraway, P.~L.~Knight, {\it Phys.~Rev.~A}
   {\bf 49}, 1266 (1994).
  \bibitem {23} P.~Goetsch, R.~Graham, F.~Haake, in preparation.
  \bibitem {24} B.~M.~Garraway, P.~L.~Knight, in {\it Quantum Optics
   VI}, ed.~by.~J.~D.~Harvey, D.~F.~Walls, Berlin, Springer 1994;\\
   B.~M.~Garraway, P.~L.~Knight, submitted to {\it Phys.~Rev.~Lett.}
  \bibitem {25} H.~J.~Carmichael, in {\it Quantum Optics VI},
   ed.~by.~J.~D.~Harvey, D.~F.~Walls, Berlin, Springer 1994.
  \bibitem {p1} A.~Barchielli, {\it Phys.~Rev.~A} {\bf 34}, 1642
   (1986).
  \bibitem {p2} A.~Barchielli, in {\it Quantum Probability and
   Applications III (Lecture Notes in Mathematics {\bf 1303})}, ed.~by
   L.~Accardi, W.~von Wadenfels, Springer, Berlin 1988, p.~37.
  \bibitem {p3} A.~Barchielli, {\it Quantum Opt.~} {\bf 2}, 423 (1990).
  \bibitem {p5} C.~W.~Gardiner, A.~S.~Parkins, P.~Zoller, {\it
   Phys.~Rev.~A} {\bf 46}, 4363 (1992).
  \bibitem {p8} R.~Loudon, {\it The Quantum Theory of Light}, Oxford
   University Press, Oxford 1983.


\end {thebibliography}
\end {document}